# Ferroelectric control of Néel vector in L10 type of antiferromagnetic films

Fanxing Zheng[1], Meng Zhu[1], Xinlu Li[1], Peina Zhang[1], Jiuzhao Liu[1], Jianting Dong[1], and Jia Zhang[1*]

[1]*School of Physics and Wuhan National High Magnetic Field Center, Huazhong University of Science and Technology, 430074 Wuhan, China*

* jiazhang@hust.edu.cn

## ABSTRACT

How to efficiently manipulate the Néel vector of antiferromagnets (AFM) by electric methods is one of the major focuses in current antiferromagnetic spintronics. In this work, we investigated the ferroelectric control of magnetism in AFM L10-MnPt/BaTiO$_3$ bilayers structures by using first-principles calculation. We studied the effect of ferroelectric polarization reversal on magnetic crystalline anisotropy (MCA) of L10-MnPt films with different interface structures. Our results predict a large perpendicular MCA in L10-MnPt films with Pt-O interface, while an in-plane MCA with Mn-O interface when they are interfaced with ferroelectric BaTiO$_3$. In addition, the magnitude of MCA for both interfaces can be modulated efficiently by the polarization reversal of BaTiO$_3$. The ferroelectric control of MCA has been analyzed based on second order perturbation theory, and it can be mainly attributed to the ferroelectric polarization driven redistribution of Pt-5$d$ orbital occupation around Fermi energy. Especially, for Mn-O interface, the Néel vector can be switched between in-plane [100] and [110] directions, or even from in-plane to out-of-plane at certain film thickness by reversing ferroelectric polarization. Our results may provide a non-volatile concept for ferroelectric control of Néel vector in L10-antiferromagnets, which could stimulate experimental investigations on magnetoelectric effect of antiferromagnets and promote its applications in low-power consumption spintronic memory devices.

# I. INTRODUCTION

Recently, electric control of antiferromagnetic states has been widely discussed and investigated[1][2][3]. Compared with ferromagnetic materials (FMs), antiferromagnetic materials (AFMs) are robust against external magnetic field, produce no stray field, and exhibit ultrafast dynamics[2][4]. Those features make AFMs appealing for applications in non-volatile memory[5][6]. However, how to efficiently manipulate the magnetic states of AFMs by electric means is one of the key issues for future development of AFMs based spintronics devices.

Currently, intensive research attention has been focused on the electric current switching of AFM states[7][8][9]. A number of theoretical and experimental works have demonstrated that in several AFMs like $Mn_2Au$, CuMnAs with specific crystal symmetry[10][11][12] and in AFM/heavy metal bilayer structures[13][14][15], the electric current can manipulate the Néel vector in AFM through spin-orbit torque (SOT) mechanism[16]. Unfortunately, in order to manipulate the antiferromagnetic state, a current density typically around $10^6$~$10^7$ A/cm$^2$ is required[9][17][18]. Therefore, a more energy efficient switching mechanism is necessary for manipulating antiferromagnetic states[19][20][21].

Instead of using electric current, an alternative promising approach to manipulate Néel vector of AFMs may be applying electric field[22][23]. In the AFMs/ferroelectrics bilayers structure, ferroelectric polarization can be reversed by applying electric field, so as to possibly manipulate the magnetic properties of adjacent AFM films. Earlier theoretical calculations have predicted magnetoelectric effect in Fe/$BaTiO_3$ interface[24]. And in FeRh/$BaTiO_3$ bilayers, it has also been proved that a moderate electric field is sufficient to adjust the AFM-FM phase transition and MCA of FeRh[25][26]. MnPt belongs to L10 type of metallic AFMs, which has high Néel temperature ($T_N$ = 970 K) and makes it suitable for practical applications. In addition, previous first-principles calculations indicate the easy axis of bulk L10-MnPt is along *c*-axis and sensitive to the number of electrons[27]. Those properties stimulate our investigations on ferroelectric control of MCA in MnPt films.

In this work, we set up L10-MnPt/BaTiO$_3$ bilayers and investigate the ferroelectric control of MCA in MnPt films with various thickness by using first-principles calculations. As shown in Figs.1(a) and 1(b), we consider two possible interface structures between MnPt and BaTiO$_3$, namely, with Pt and Mn layers in MnPt sit on top of O atoms in TiO$_2$ plane of BaTiO$_3$. There is a small in-plane lattice mismatch (0.28%) between L10-MnPt (*a*=*b*=3.980 Å, *c*=3.720 Å)[28] and BaTiO$_3$ (*a*=*b*=3.991 Å, *c*=4.035 Å)[29], which suggests the proposed L10-MnPt/BaTiO$_3$ bilayer structures could be experimentally achievable. The calculation results demonstrate that for Pt-O interface, MnPt films have a large perpendicular magnetic anisotropy (PMA) high up to several mJ/m$^2$ (Hereafter the MCA is defined as the total MCA energy of film in the unit cell divided by the unit cell area which means the MCA of the films per unit area.), while for Mn-O interface, the magnetic easy axis of MnPt keeps in-plane in the studied film thickness. Importantly, our results prove the concept that, instead of electric current, the reversal of ferroelectric polarization could remarkably modify the magnitude of MCA in adjacent AFM thin films. What's more, for Mn-O interface case, the Néel vector orientations of MnPt films can be realized by reversing ferroelectric polarization of BaTiO$_3$.

## II. CALCULATION METHODS

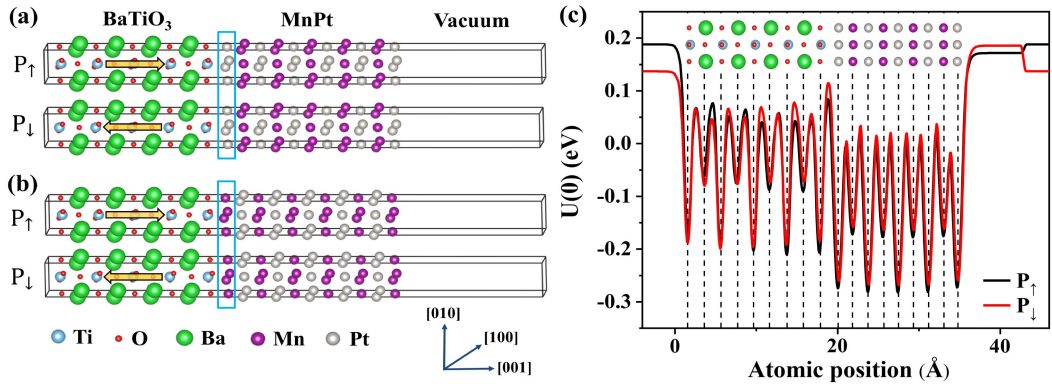

FIG. 1. The side view of MnPt (4 u.c.)/BaTiO$_3$ (4 u.c.)/vacuum structures with Pt-O (a) and Mn-O (b) interfaces for P$_\uparrow$ and P$_\downarrow$ ferroelectric polarization states. The light blue squares highlight the interfacial layers of L10-MnPt films. Yellow arrows indicate the directions of ferroelectric polarization. The crystal coordinates axes are also indicated in the bottom right corner. (c) The planar averaged electrostatic potential energy distribution in MnPt (4 u.c.)/BaTiO$_3$ (4 u.c.)/vacuum structure for P$_\uparrow$ (black line) and P$_\downarrow$ (red line) polarization states.

The first-principles calculations are performed by employing the projector augmented wave (PAW) pseudopotential[30] and the Perdew-Burke-Ernzerhof (PBE) type of generalized gradient approximation (GGA) for the exchange correlation potential[31] as implemented in the Vienna ab initio simulation package (VASP)[32]. The L1$_0$-MnPt/BaTiO$_3$ bilayers are modeled by using a supercell with the vacuum thickness of no less than 15 Å as it is shown in Fig.1(a). The thickness of MnPt films is varied from one to five unit cells (1-5 u.c.), while the thickness of BaTiO$_3$ is fixed to be 4 u. c. which is proved to be sufficient for establishing its ferroelectric stability[33]. The in-plane lattice constant of the supercell is fixed to be the bulk value of BaTiO$_3$ (3.991Å). The dipole correction is applied in order to minimize the artificial electrostatic interactions between asymmetric surface layers[34]. As shown in Fig. 1(c), the step of electrostatic potential appears in vacuum region due to dipole correction, which eliminates the artificial electric field across the neighbouring supercell.

An energy cutoff of 500 eV and $12 \times 12 \times 1$ $k$ point mesh in the Brillouin zone are used for structure relaxation until the force on each atom is less than 10 meV /Å. During the relaxations of ionic positions, MnPt atoms are allowed to fully relax, while only one unit cell of BaTiO$_3$ on the interface is allowed to relax, and the other three unit cells of BaTiO$_3$ away from MnPt-BaTiO$_3$ interface are fixed to their bulk positions. The P$_\uparrow$ and P$_\downarrow$ represent the Ti atoms displacement toward and away from MnPt interface as shown in Fig.1(a). By calculating the interface binding energy, Mn-O interface may be more stable in comparison to Pt-O interface (See Appendix A). However, experimentally, both interface structures might be fabricated by depositing monolayer Pt or Mn before preparing MnPt films. The interface magnetic states (FM or AFM) have also been investigated in Appendix B, which indicate the reversal of the ferroelectric polarization does not change the interface AFM order both for Mn-O and Pt-O interface structures.

The MCA is calculated based on the force theorem. First, a self-consistent calculation in the absence of spin-orbit coupling (SOC) is performed to obtain charge density by using $24 \times 24 \times 1$ $k$ points. Then, the MCA is evaluated by using a much denser $k$-point mesh of $35 \times 35 \times 1$ which has been checked for convergence. The MCA is obtained by taking the band energy difference for magnetization along in-plane and out-of-plane directions as: MCA= $E_{band}$[abc]-$E_{band}$[001], where [abc] indicates the Néel vector is along in-plane [100] or [110] directions.

The Néel vector orientations of MnPt films are determined by the total magnetic anisotropy energy(MAE=MCA+$M_{dd}$) which includes MCA as well as the magnetic dipole-dipole anisotropy energy ($M_{dd}$). However, $M_{dd}$ originating from magnetostatic interaction of magnetic moments has not been included in the first-principles calculations[35]. We explicitly calculate $M_{dd}$ of MnPt films by employing real-space summation as described in Appendix C.

## III. RESULTS AND DISCUSSIONS

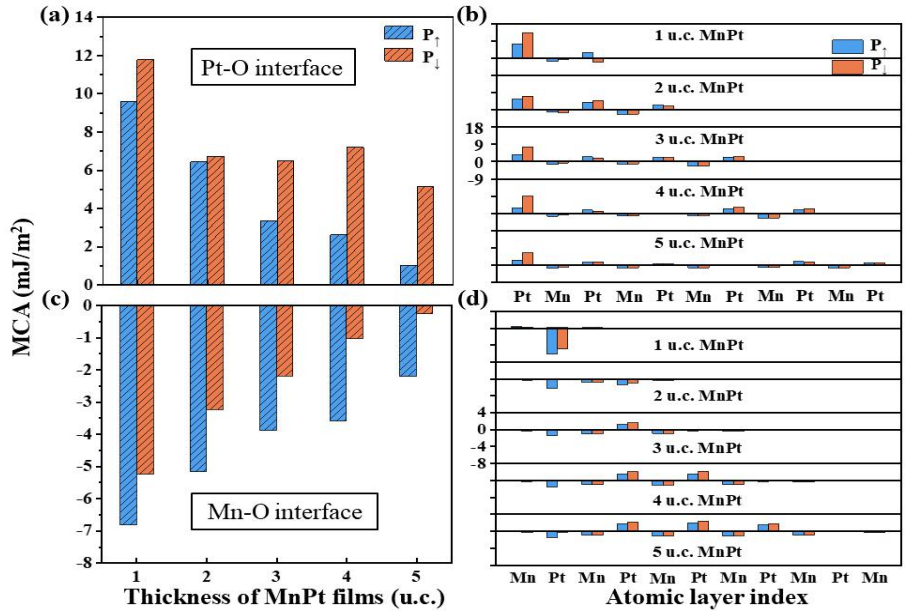

FIG. 2. Thickness dependence of MCA(=$E_{band}$[100]-$E_{band}$[001]) in MnPt/BaTiO$_3$ bilayers for $P_\uparrow$ (blue) and $P_\downarrow$ (orange) states for the cases of (a) Pt-O and (c) Mn-O interfaces. The layer-resolved MCA for the cases of (b) Pt-O and (d) Mn-O interfaces. Figures in (b) and (d) have the same tick labels for MCA, and for simplicity, only tick labels for 3 u.c. MnPt films

have been shown.

As shown in Figs. 2(a) and 2(c), one can observe that the MCA is large and typically around several mJ/m$^2$ in MnPt films by interfacing with BaTiO$_3$. What's more, the reversal of the ferroelectric polarization can regulate the magnitude of MCA by several times, which demonstrates the efficient ferroelectric control on magnetism of AFM MnPt films. Remarkably, for the case of Pt-O interface, MnPt films significantly enhance perpendicular magnetic anisotropy (PMA) in comparison with the bulk L10-MnPt (MCA=0.176 meV/u.c., easy axis is along *c*-axis)[36] and the P$_↓$ polarization yields a larger PMA than P$_↑$ polarization state. On the other hand, for the case of Mn-O interface, MCA is always negative for the investigated MnPt films under both polarization states of BaTiO$_3$, and the magnitude of MCA decreases monotonically as the thickness of MnPt layers increases [Fig. 2(c)]. In some simple cases, the MCA of films could be described by a model as: $K=K_{is}+K_b*t$, where $K$ represents the total MCA of films, $K_{is}$ represents the interface and surface MCA contributions and $K_b$ is the bulk MCA, $t$ is the film thickness. Ideally, the MCA may show a linear dependence on the film thickness with a slope as the bulk MCA and the intercept as the surface/interface MCA contribution. However, this asymptotic behavior is not present in the studied MnPt films, possibly due to complicated thickness dependence of MCA on surface/interface contribution, or formation of quantum well states.

To gain more insight on MCA and the ferroelectric control of MCA, we could evaluate the layer resolved MCA by projecting the SOC energy ($E_{SOC}$) on each atom and taking the energy difference when the magnetization is along [100] and [001] directions. The MCA contribution thus can be reversed on each atomic layer as[37]:

$$MCA^i \approx \frac{1}{2}(E^i_{SOC}[100] - E^i_{SOC}[001]) \qquad (1)$$

where *i* represents atomic layer index and a positive $MCA^i$ means that atomic layer contributes positive MCA and tends to PMA.

For Pt-O interface case, as it is shown in Fig. 2(b), the Pt layers contribute positive MCA and the interfacial Pt atomic layer contributes the largest MCA. The ferroelectric polarization reversal of BaTiO$_3$ significantly modifies the MCA on interfacial Pt layer and therefore leads to ferroelectric control of total MCA in MnPt films. The mechanism for PMA (perpendicular magnetic anisotropy) and the ferroelectric control of MCA mechanism for Pt-O interface case can be attributed to orbital hybridization between Pt-5$d$ and O-2$p$ orbitals as we will discuss later. The crucial role of the interfacial Pt atoms on MCA and the polarization control of MCA is further revealed from the Figs. 2(b) and 2(d), which is owing to the inherently large SOC constant ξ and induced magnetic moment on Pt (the magnetic moment on Pt is around 0.1 μ$_B$ as it is shown in Appendix D). And similar significant MCA contribution from nonmagnetic atom (or induced magnetic moment on nonmagnetic atoms) has been reported in previous literatures[38][39][40][41].

For the case of Mn-O interface, as it is shown in Fig. 2(d), the Pt atoms next to Mn-O interface exhibit negative contribution to MCA and also show the modulation upon polarization reversal. As the MnPt thickness increases, the Pt atoms in the middle layer of MnPt films start to restore the bulk like feature, and start to contribute positive MCA. Consequently, the magnitude of negative MCA decreases monotonically as the thickness of MnPt layers increases as it is shown in Fig. 2(c). The in-plane MCA and its ferroelectric control of Néel vector can also be attributed to the change of electronic structure on Pt atomic layer nearest to Mn-O interface as we will discuss later.

In the following, we will qualitatively analyze the origin of MCA and its ferroelectric polarization controlled effect in MnPt/BaTiO$_3$ bilayers based on the second order perturbation theory[42][43][44]. In perturbation theory, the spin-orbit coupling (SOC) Hamiltonian can be written as:

$$H_{soc} = \xi(r)\vec{\sigma} \cdot \vec{L} \quad (2)$$

where $\vec{L}$ is the orbital angular momentum operator, $\vec{\sigma}$ is the Pauli matrix, and $\xi(r)$ is the atomic spin-orbit coupling strength. Generally, since SOC energy is at the order of tens of meV, it can be treated as energy perturbation. The second order energy correction due to SOC is related to the magnetization direction $M_\eta$ and can be written as[42][43][44][45]:

$$\Delta E_{\text{soc}}(M_\eta) = \frac{\xi^2}{4} \sum_{\alpha,\beta} \sum_{o,u} \frac{\left|\left\langle \psi_o^\alpha \left| (\vec{\sigma} \cdot \vec{L})_{M_\eta}^{\alpha\beta} \right| \psi_u^\beta \right\rangle\right|^2}{\varepsilon_o^\alpha - \varepsilon_u^\beta} \quad (3)$$

where $\alpha$, $\beta$ are the spin indexes ($\alpha$, $\beta = \uparrow, \downarrow$), $\psi_o^\alpha$ and $\psi_u^\beta$ are the occupied (above Fermi energy) and unoccupied (below Fermi energy) wave functions in the absence of SOC, $\varepsilon_o^\alpha$ and $\varepsilon_u^\beta$ are the corresponding band eigenvalues.

The SOC Hamiltonian is shown in eq.(2), and therefore, the energy correction $\Delta E_{\text{soc}}(M_\eta)$ in eq.(3) is related to the magnetization orientation ($M_\eta$), and in consequence, leads to MCA. By integrating over $k$ in Brillouin Zone, $\Delta E_{\text{soc}}(M_\eta)$ can be approximately expressed by the product of the density of states (DOS) and the matrix elements of $\vec{\sigma} \cdot \vec{L}$ between a pair of $d$ orbitals as follows[37][45]:

$$\Delta E_{\text{SOC}}(M_\eta) = \frac{\xi^2}{4} \sum_{\mu,\mu'} P_{M_\eta}^{\alpha\beta}(d_\mu, d_{\mu'}) \int_{-\infty}^{\varepsilon_F} d\varepsilon \int_{\varepsilon_F}^{\infty} d\varepsilon' \frac{\rho_\mu^\alpha(\varepsilon)\rho_{\mu'}^\beta(\varepsilon')}{\varepsilon - \varepsilon'};$$

$$P_{M_\eta}^{\alpha\beta}(d_\mu, d_{\mu'}) = \left|\left\langle d_\mu \left| (\vec{\sigma} \cdot \vec{L})_{M_\eta}^{\alpha\beta} \right| d_{\mu'} \right\rangle\right|^2 \quad (4)$$

where $d_\mu$ belongs to one of the five $d$-orbitals $d_{xy}$, $d_{yz}$, $d_{xz}$, $d_{x^2-y^2}$ and $d_{z^2}$. $\rho_\mu^\alpha(\varepsilon)$ is the density of states (DOS) for spin $\alpha$ and $\varepsilon_F$ is the Fermi energy. The non-vanishing matrix element $(\vec{\sigma} \cdot \vec{L})_{M_\eta}^{\alpha\beta}$ between a pair of $d$-orbitals will contribute to the SOC energy correction and favor the magnetization along $M_\eta$ direction. From the above perturbation analysis, one can realize that the MCA will be closely related to the non-vanishing matrix element $(\vec{\sigma} \cdot \vec{L})_{M_\eta}^{\alpha\beta}$ and the involved DOS distribution around Fermi energy.

Hereafter, we will first analyze the MCA and its polarization regulation mechanism for Pt-O case by taking MnPt (4 u.c.)/BaTiO$_3$ as an example. Figs. 3(a) and 3(b) show the $d$ orbitals resolved MCA on interfacial Pt atomic layer for two polarization states in MnPt (4 u.c.)/BaTiO$_3$ bilayers. It clearly indicates that the orbital pair of $(d_{z^2}, d_{yz})$ contributes the largest positive MCA and such contribution has been significantly modulated upon ferroelectric polarization reversal from $P_\uparrow$ to $P_\downarrow$ state. By considering the fact that, the Pt-O interface distance has been changed from 2.20 Å to 2.28 Å by polarization reversal from $P_\uparrow$ to $P_\downarrow$, one can infer the control of MCA may be correlated to the modification of interface orbital hybridization.

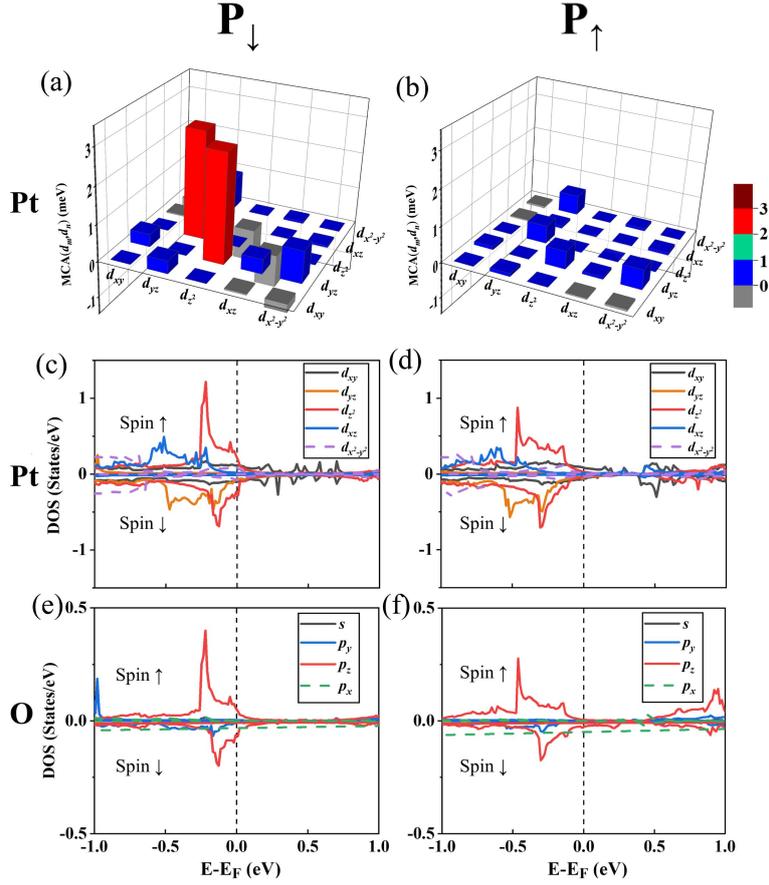

FIG. 3. The $d$-orbital resolved MCA on interfacial Pt atomic layer for $P_\downarrow$ (a) and $P_\uparrow$ (b) ferroelectric polarization states for the case of Pt-O interface in MnPt (4 u.c.)/BaTiO$_3$ bilayer. The orbital-resolved DOS on interfacial Pt and O atoms in MnPt (4 u.c.)/BaTiO$_3$ bilayers for $P_\downarrow$ and $P_\uparrow$. (c) and (d) for Pt-5$d$ orbitals; (e) and (f) for O-2$p$ orbitals. The Fermi level is set to zero.

Indeed, Figs. 3(c)-3(f) show the density of states (DOS) for Pt-5$d$ and O-2$p$ orbitals on Pt-O interface, which clearly indicate the orbital hybridization and the formation of bonding states between interfacial Pt and O atoms. This is evident from

Figs. 3(c) and 3(e) that the interfacial Pt-$d_{z^2}$ and O-$p_z$ peaks have similar shapes and are located at the same position below Fermi energy. (Please note that the DOS of $d_{xz}$ and $d_{yz}$ orbitals do not degenerate on interfacial Pt atom due to the *x, y* symmetry breaking at the interface.) On the other hand, based on the second order perturbation theory[37], for magnetic moment along $M_z$ direction ([001]), the non-vanishing matrix elements involved $(d_{z^2}, d_{yz})$ orbital pair are $\langle d_{yz} | (\vec{\sigma} \cdot \vec{L})_{M_z}^{\uparrow\downarrow} | d_{z^2} \rangle = -i\sqrt{3}$, while for magnetic moment along in-plane $M_x$ direction, such spin-flipped SOC matrix element is missing. In consequence, the presence of $d_{z^2}$ orbital around Fermi energy is essentially crucial for contributing to SOC energy correction $\Delta E_{soc}(M_z)$ and thus favors positive MCA and resultant PMA.

The reduction of MCA from $P_\downarrow$ to $P_\uparrow$ polarization reversal can also be understood from the modification of orbital-resolved DOS, which is shown in Figs. 3(c)-3(f). Notably, the interfacial Pt-5*d* (mostly $d_{z^2}$ and $d_{yz}$) and O-2*p* (mostly $p_z$ and $p_y$) bonding states move simultaneously downward across the Fermi level through polarization reversal from $P_\downarrow$ to $P_\uparrow$. For $P_\downarrow$ polarization state, the interface hybridization between the Pt-5*d* and O-2*p* orbitals leads to a larger $d_{yz}$ and $d_{z^2}$ DOS presented around Fermi energy than the case of $P_\uparrow$ polarization. Therefore, a larger contribution to positive MCA contributed by $(d_{z^2}, d_{yz})$ orbital pair for $P_\downarrow$ polarization. In order to further confirm the ferroelectric polarization reversal effect on interfacial Pt layer, we also calculate the orbital projected band structure along high-symmetry lines as it is shown in Figs. 4(a) and 4(b). It is clear that when the ferroelectric polarization is reversed from $P_\downarrow$ to $P_\uparrow$, the $d_{z^2}$ orbital contribution to energy bands around Γ points on interfacial Pt layer has been shifted below Fermi level.

By taking magnetic dipole-dipole anisotropy energy ($M_{dd}$) into account, the total MAE as a function of MnPt film thickness for Pt-O interface is shown in Fig.C1 in

Appendix C. One can find that $M_{dd}$ is positive and increase linearly for film thickness from 1 to 5 u.c. and suggest the magnetic dipole-dipole energy favors out-of-plane easy axis for AFM MnPt films. Therefore, in addition to MCA, the MAE ensures the Néel vector of MnPt films with Pt-O interface always points along out-of-plane [001] direction. The magnitude of MAE can be manipulated by several times through ferroelectric polarization reversal, which demonstrates the desired efficient ferroelectric control of MAE for AFM films.

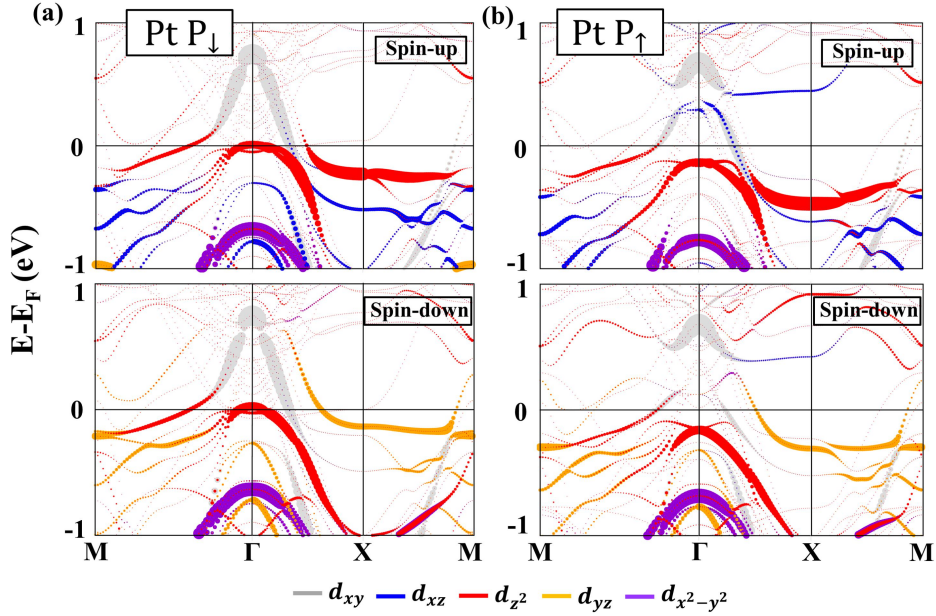

FIG. 4. The *d*-orbital (upper panels for spin up; lower panels for spin down) projected band structures on interfacial Pt atomic layer for $P_\downarrow$ (a) and $P_\uparrow$ (b) ferroelectric polarization states in MnPt (4u.c.)/BaTiO$_3$ bilayers. The size of symbols represents the projected weight of different *d*-orbitals. The Fermi level is set to zero.

In the following, we will qualitatively analyze the MCA and its polarization-controlled effect for the case of Mn-O interface. As previously shown in Fig. 2(c), for Mn-O interface, the MCA of MnPt films favor in-plane magnetic easy axis for both ferroelectric polarization states. This result can also be understood from the electronic structure of Pt atomic layer nearest to Mn-O interface. Figs. 5(a) and (b) show the *d*-orbital resolved MCA on the Pt layer nearest to Mn-O interface in MnPt (4 u.c.)/BaTiO$_3$ bilayers. It is clear that now the orbital pairs of $(d_{x^2-y^2}, d_{xy})$ and $(d_{xz}, d_{yz})$ contribute prominent negative MCA (favor in-plane easy axis). Based on the second order perturbation theory, for magnetic moment along $M_x$ direction ([100]), the nonzero matrix elements involved $(d_{x^2-y^2}, d_{xy})$ and $(d_{xz}, d_{yz})$ orbital pairs are the spin-flipped terms $\langle d_{x^2-y^2} | (\vec{\sigma} \cdot \vec{L})_{M_x}^{\uparrow\downarrow} | d_{xy} \rangle = -2i$ and $\langle d_{xz} | (\vec{\sigma} \cdot \vec{L})_{M_x}^{\uparrow\downarrow} | d_{yz} \rangle = -i$, while for magnetic moment along $M_z$ direction, those matrix elements are vanishing. Therefore, the presence of $d_{x^2-y^2}$, $d_{xy}$, $d_{yz}$ and $d_{xz}$ orbitals around Fermi energy as it is shown in Fig. 5(c-d) could be partly responsible for negative MCA. On the other hand, in sharp contrast with Pt-O interface, the missing of interface bonding between Pt-$5d$ and O-$2p$ orbital for Mn-O interface case would reduce the positive MCA contribution by $(d_{z^2}, d_{yz})$ orbital pair. In consequence, the MnPt films with Mn-O interface result in in-plane magnetic easy axis. As shown in Figs. 5(a) and 5(b), by reversing the polarization from $P_\downarrow$ to $P_\uparrow$, the main change of MCA on Pt layer nearest to Mn-O interface is the $(d_{yz}, d_{x^2-y^2})$ orbital pair which has been changed from positive to negative MCA contribution.

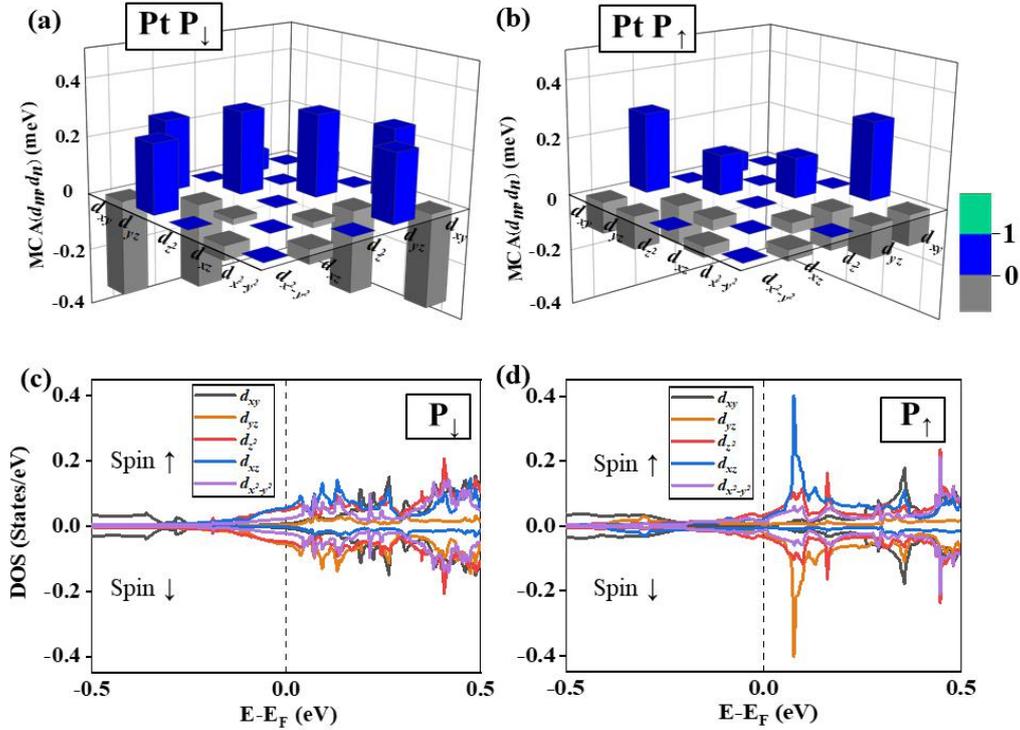

FIG. 5. The $d$-orbital resolved MCA on Pt atomic layer nearest to Mn-O interface for $P_\downarrow$ (a)

and $P_\uparrow$ (b) ferroelectric polarization states for the case of Mn-O interface in MnPt (4 u.c.)/BaTiO$_3$ bilayer. And the corresponding $d$-orbital resolved density of states for $P_\downarrow$ (c) and $P_\uparrow$ (d) polarization states. The Fermi level is set to zero.

The obvious redistribution of $d$-orbital occupation around Fermi energy occurs due to the polarization reversal as shown in Figs. 5(c) and 5(d). One can observe that the DOS around Fermi energy mainly include $d_{z^2}$, $d_{x^2-y^2}$, $d_{yz}$ and $d_{xz}$ orbitals and the magnitude of DOS for such $d$-orbitals decreases by reversing the ferroelectric polarization state from $P_\downarrow$ to $P_\uparrow$. There are similar $d$-orbitals pairs of nonzero matrix elements for SOC Hamiltonian $H_{SOC}$ existing for magnetic moment (i.e. Néel vector) along [100] and [110] directions[37]. And for such orbital pairs, the value of SOC matrix element is different for those two in-plane magnetization orientations. For instance, for the orbital pair of $(d_{xz}, d_{x^2-y^2})$, for [100] case the non-vanishing SOC matrix element is $\langle d_{xz} | (\vec{\sigma} \cdot \vec{L})^{\uparrow\downarrow}_{M_x} | d_{x^2-y^2} \rangle = -1$, while for magnetic moment along [110] direction it is $\langle d_{xz} | (\vec{\sigma} \cdot \vec{L})^{\uparrow\downarrow}_{M_{[110]}} | d_{x^2-y^2} \rangle = -\frac{\sqrt{2}}{2}$. Therefore, the reversal of ferroelectric polarization will modify the electronic structures and DOS around Fermi energy, thus change the relative SOC energy contribution for such orbital pairs for magnetic moment along in-plane [100] and [110] directions. In consequence, it leads to the possible ferroelectric polarization driven switching between in-plane easy axis as we discuss below.

By considering the total magnetic anisotropy contributed from MCA and magnetic dipole-dipole anisotropy energy $M_{dd}$, Fig. 6(a) depicts the calculated MAE as a function of MnPt films thickness in MnPt/BaTiO$_3$ bilayers with Mn-O interface. One can observe that by flipping the ferroelectric polarization from $P_\uparrow$ to $P_\downarrow$, the sign of MAE will be changed from negative to positive for 5 u.c. MnPt film, which indicates the Néel vector of MnPt films will be re-orientated from in-plane to out-of-plane direction. When the thickness of MnPt films is less than 5 u.c., the Néel vector of MnPt keeps in-plane. Fig. 6(b) shows the MAE difference

(ΔMAE=MAE[100]-MAE[110], the positive ΔMAE means the Néel vector of MnPt favors [110] direction, and vice versa) between in-plane [100] and [110] directions. The calculation results indicate that the Néel vector of MnPt film can be switched between in-plane [100] or [110] directions. For instance, for 1-2 u.c. MnPt films, by reversing the ferroelectric polarization from $P_\uparrow$ to $P_\downarrow$, the Néel vector will switch from [100] to [110] direction, while for 3 u.c. MnPt films the same polarization reversal will re-orientate Néel vector from [110] to [100] direction. The above discussed ferroelectric switching of Néel vector of AFM films is a desired non-volatile method for controlling AFM states by electric field.

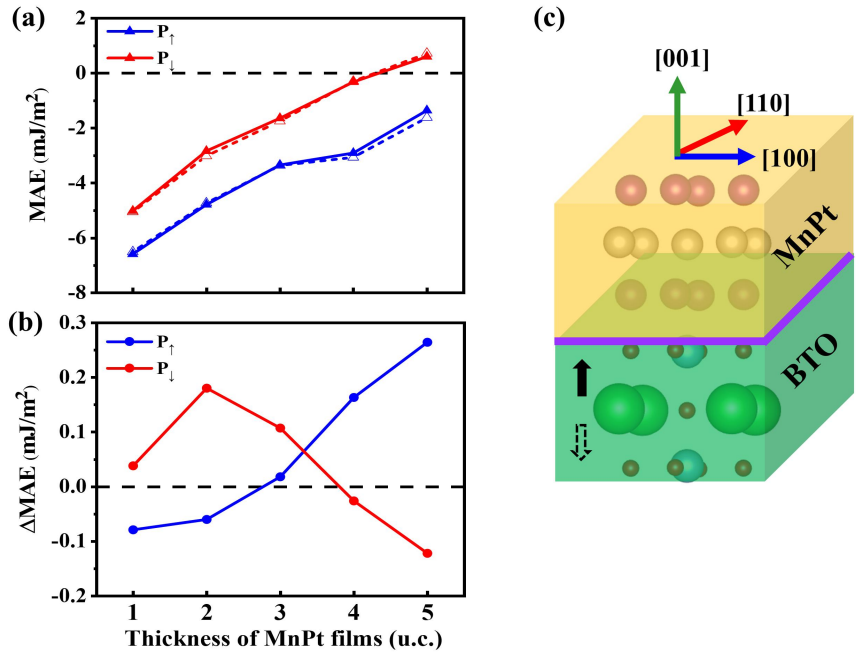

FIG. 6. (a) The calculated MAE(=E[abc]-E[001], where [abc]=[100] or [110]. The solid and dashed lines indicate MAE[100] and MAE[110], respectively.) and (b) ΔMAE(=MAE[100]-MAE[110]) as a function of MnPt films thickness in MnPt/BaTiO$_3$ bilayer for Mn-O interface. The red and blue lines represent the MAE for $P_\downarrow$ and $P_\uparrow$ polarization states. (c)The illustration of ferroelectric control of Néel vector in MnPt films. The black arrows indicate the polarization direction of BaTiO$_3$ (BTO), and the blue, red and green arrows represent Néel vector along in-plane [100], [110] or out-of-plane [001] directions, respectively.

In recent experiments, a large electric current density is needed to switch the easy axis between [110] and [100] directions in Mn$_2$Au films[12][19][20]. Here, we demonstrate that for designed AFM MnPt films, the manipulation of Néel vector can be realized by polarization reversal of adjacent ferroelectric materials. Such manipulation on AFM state is non-volatile and more energy efficient in comparison with current-driven switching. The ferroelectric manipulation of AFM states should not be limited to the representative material MnPt films as we studied in this work, and this phenomenon should be quite general for other AFM/ferroelectric materials bilayer structures.

## IV. SUMMARY

In summary, ferroelectric control of MCA on L10 type of antiferromagnets MnPt films has been investigated by using first-principles calculations. We consider two possible interface configurations with BaTiO$_3$, namely Pt-O and Mn-O interfaces. For the case of Pt-O interface, there are giant PMA induced by the ferroelectric BaTiO$_3$ due to the rising of $d_{z^2}$ orbital by Pt-5$d$ and O-2$p$ interface bonding, while the MnPt films have in-plane MCA for Mn-O case in the studied thickness rang (1-5 u.c.). Interestingly, for Mn-O interface, at certain thickness of MnPt films, the Néel vector can be switched between in-plane [100] or [110] directions, or even re-oriented from in-plane to out-of-plane direction by reversing the ferroelectric polarization of BaTiO$_3$. These calculation results may provide a concept of non-volatile electric control of Néel vector in antiferromagnets, which should be appealing for the development of high-speed, high-density and low-power consumption antiferromagnets based memory devices.

## ACKNOWLEDGEMENT

Jia Zhang acknowledges support from the National Natural Science Foundation of China (grants No. 11704135). Computations are partly performed on the Platform for Data-Driven Computational Materials Discovery at the Songshan Lake Materials Laboratory, Dongguan, China. The authors thank Prof. Xiaohui Liu from School of Physics, Shandong University for helpful discussions.

# APPENDIX A: Binding energy with different interfaces for MnPt/BaTiO$_3$ bilayers.

To gain insight into interface stability, we calculated the binding energy with Pt-O and Mn-O interfaces by taking MnPt (4 u.c.)/BaTiO$_3$ as an example (see Table I). From the calculation results, we can find the interface binding energy of Mn-O interface is larger than Pt-O interface both for P$_\downarrow$ and P$_\uparrow$ polarization states which suggests Mn-O interface may be more energy stable in comparison to Pt-O interface. Regarding the surface stability, as we notice that the modulation and contribution of MCA in MnPt/BaTiO$_3$ largely originates from the interface MnPt layers, instead of surface MnPt layers. In consequence, one would expect similar conclusions by considering other possible surface terminations.

Table I. The interface binding energy for Pt-O and Mn-O interface configurations, where $E_0$ (interface binding energy) = $E_0$ (MnPt/BaTiO$_3$) - $E_0$ (MnPt) - $E_0$ (BaTiO$_3$).

| Interface Structure | Polarization direction | $E_0$(MnPt/BaTiO$_3$) (eV) | $E_0$(MnPt) (eV) | $E_0$(BaTiO$_3$) (eV) | Interface binding energy(eV) |
|---|---|---|---|---|---|
| Mn-O | P↑ | -328.84293 | -142.47310 | -184.181108 | -2.189 |
|  | P↓ | -328.71805 | -142.47844 | -183.637820 | -2.602 |
| Pt-O | P↑ | -323.44153 | -138.22084 | -183.980770 | -1.240 |
|  | P↓ | -323.12561 | -138.24023 | -183.823240 | -1.062 |

# APPENDIX B: Stable magnetic states (FM or AFM) at MnPt/BaTiO$_3$ interfaces.

We further analyzed the stable interface magnetic states (FM or AFM) by taking the MnPt (4 u.c.)/BaTiO$_3$ as an example. We calculate the total energy of the bilayers by setting the interface MnPt layer as FM or AFM states (see Table II). The results indicate that the reversal of the polarization direction does not change the interfacial

AFM states both for Mn-O and Pt-O interfaces structure.

Table II. Total energy for FM and AFM interface magnetic configurations. The energy differences between FM and AFM states are defined as $\Delta E = E_0 (AFM) - E_0 (FM)$.

| Interface Structure | Polarization direction | $E_0$(AFM) (eV) | $E_0$(FM) (eV) | $\Delta E$ (eV) | Stable interface magnetic configurations |
|---|---|---|---|---|---|
| Mn-O | P↑ | -328.84293 | -328.30910 | -0.534 | AFM |
| Mn-O | P↓ | -328.71805 | -328.11157 | -0.606 | AFM |
| Pt-O | P↑ | -323.44153 | -322.86483 | -0.577 | AFM |
| Pt-O | P↓ | -323.12561 | -322.59067 | -0.535 | AFM |

## APPENDIX C: Magnetic dipole-dipole energy contribution to MAE.

The MAE includes contributions from MCA and magnetic dipole-dipole anisotropy energy ($M_{dd}$)[41][46]. We calculate $M_{dd}$ (defined as the magnetic dipole energy difference by taking [001] as reference: $M_{dd}=E_{dd}[abc]-E_{dd}[001]$, where [abc] indicates the direction of Néel vector). The Néel vector dependant magnetic dipole energy $E_{dd}$ is calculated based on the following equation by direct summation in real space with cut-off radius of 800 Å[47]:

$$E_{dd} = \frac{\mu_0}{4\pi} \sum_{i<j} \frac{1}{r_{ij}^3}\left[\boldsymbol{m_i} \cdot \boldsymbol{m_j} - 3\frac{(\boldsymbol{r_{ij}} \cdot \boldsymbol{m_i})(\boldsymbol{r_{ij}} \cdot \boldsymbol{m_j})}{r_{ij}^2}\right]$$

where $\boldsymbol{m_{i,j}}$ are the magnetic moments on atomic sites $i$ and $j$, $\boldsymbol{r}_{ij}$ is the vector pointing from $i$ to $j$, and $\mu_0$ is the permeability in vacuum.

The convergence criterion of $M_{dd}$ was set to be $10^{-10}$ meV. Fig. C1 depicts the MCA and the magnetic dipole-dipole anisotropy $M_{dd}$ to MAE as a function of films thickness with Pt-O interface. One can observe that, the magnetic dipole-dipole anisotropy energy $M_{dd}$ depends linearly on film thickness both for $P_↓$ and $P_↑$ polarization states with the value from 0.12 to 0.71 mJ/m$^2$. The $M_{dd}$ as a function of thickness for Mn-O interface shows similar behavior.

FIG.C1. MnPt thickness dependence of MCA, magnetic dipole-dipole anisotropy energy $M_{dd}$ and the total MAE of MnPt/BaTiO$_3$ bilayers with Pt-O interface.

## APPENDIX D. Magnetic moments on Mn and Pt atoms.

By taking the MnPt (2 u.c.)/BaTiO$_3$ and MnPt (4 u.c.)/BaTiO$_3$ as an example, we list magnetic moments of Mn and Pt atoms with two different interface structures in table III. The induced magnetic moments on Pt atoms typically are around 0.1 $\mu_B$.

Table III. The magnetic moments of Mn atoms and the magnetically induced Pt atoms ($\mu_B$) in the Mn-O interface and Pt-O interface configurations. Please note that for each layer there are two atoms.

| Mn-O Interface | | Interface Layer | | Middle Layers | | | | | Surface Layer | |
|---|---|---|---|---|---|---|---|---|---|---|
| | | Mn | Pt | Mn | Pt | Mn | Pt | Mn | Pt | Mn |
| 2u.c. MnPt BaTiO$_3$ | P↑ | -3.355 | -0.058 | — | — | -3.728 | — | — | -0.072 | -3.944 |
| | | 3.340 | 0.084 | — | — | 3.703 | — | — | 0.067 | 3.933 |
| | P↓ | -3.649 | -0.089 | — | — | -3.733 | — | — | -0.069 | -3.942 |
| | | 3.571 | 0.062 | — | — | 3.703 | — | — | 0.069 | 3.945 |

| | | Pt | Mn | Pt | Mn | Pt | Mn | Pt | Mn | Pt |
|---|---|---|---|---|---|---|---|---|---|---|
| 4u.c. MnPt / BaTiO$_3$ | P↑ | -3.354 | -0.060 | -3.706 | -0.102 | -3.705 | -0.100 | -3.717 | -0.069 | -3.926 |
| | | 3.441 | 0.087 | 3.681 | 0.099 | 3.704 | 0.101 | 3.720 | 0.074 | 3.932 |
| | P↓ | -3.671 | -0.102 | -3.704 | -0.105 | -3.709 | -0.100 | -3.719 | -0.069 | -3.928 |
| | | 3.569 | 0.059 | 3.671 | 0.099 | 3.708 | 0.100 | 3.722 | 0.073 | 3.934 |

| Pt-O Interface | | Interface Layer | | Middle Layers | | | | | Surface Layer | |
|---|---|---|---|---|---|---|---|---|---|---|
| | | Pt | Mn | Pt | Mn | Pt | Mn | Pt | Mn | Pt |
| 2u.c. MnPt / BaTiO$_3$ | P↑ | -0.091 | -3.654 | — | — | -0.111 | — | — | -3.652 | -0.102 |
| | | 0.093 | 3.683 | | | 0.120 | | | 3.662 | 0.115 |
| | P↓ | -0.087 | -3.636 | — | — | -0.107 | — | — | -3.636 | -0.111 |
| | | 0.102 | 3.663 | | | 0.126 | | | 3.663 | 0.109 |
| 4u.c. MnPt / BaTiO$_3$ | P↑ | -0.096 | -3.644 | -0.1 | -3.703 | -0.098 | -3.714 | -0.109 | -3.658 | -0.108 |
| | | 0.090 | 3.663 | 0.117 | 3.723 | 0.104 | 3.711 | 0.109 | 3.655 | 0.106 |
| | P↓ | -0.085 | -3.607 | -0.098 | -3.713 | -0.094 | -3.711 | -0.107 | -3.659 | -0.108 |
| | | 0.104 | 3.663 | 0.117 | 3.723 | 0.106 | 3.712 | 0.110 | 3.657 | 0.106 |

**REFERENCE:**

[1]. V. Baltz, A. Manchon, M. Tsoi, T. Moriyama, T. Ono, and Y. Tserkovnyak, Rev. Mod. Phys. **90**, 015005 (2018).

[2]. T. Jungwirth, X. Marti, P. Wadley, and J. Wunderlich, Nat. Nanotechnol. **11**, 231 (2016).

[3]. T. Jungwirth, J. Sinova, A. Manchon, X. Marti, J. Wunderlich, and C. Felser, Nat. Phys. **14**, 200 (2018).

[4]. T. Shiino, S. H. Oh, P. M. Haney, S. W. Lee, G. Go, B. G. Park, and K. J. Lee, Phys. Rev. Lett. **117**, 087203 (2016).

[5]. X. Marti, I. Fina, C. Frontera, J. Liu, P. Wadley, Q. He, R. J. Paull, J. D. Clarkson, J. Kudrnovsk, I. Turek, J. Kune, D. Yi, J. H. Chu, C. T. Nelson, L. You, E. Arenholz, S. Salahuddin, J. Fontcuberta, T. Jungwirth, and R. Ramesh. Nat. Mater. **13**, 367 (2014).

[6]. P. Wadley, B. Howells, J. Železný, C. Andrews, V. Hills, R. P. Campion, V. Novák, K. Olejník, F. Maccherozzi, S. S. Dhesi, S. Y. Martin, T. Wagner, J. Wunderlich, F. Freimuth, Y. Mokrousov, J. Kuneš, and J. S. Chauhan, Science **351**, 587 (2016).

[7]. A. Churikova, D. Bono, B. Neltner, A. Wittmann, L. Scipioni, A. Shepard, T. Newhouse-Illige, J. Greer, and G. S. D. Beach, Appl. Phys. Lett. **116**, 022410 (2020).

[8]. C. C. Chiang, S. Y. Huang, D. Qu, P. H. Wu, and C. L. Chien, Phys. Rev. Lett. **123**, 227203 (2019).

[9]. T. Hajiri, S. Ishino, K. Matsuura, and H. Asano, Appl. Phys. Lett. **115**, 052403 (2019).

[10]. N. Bhattacharjee, A. A. Sapozhnik, S. Y. Bodnar, V. Y. Grigorev, S. Y. Agustsson, J. Cao, D. Dominko, M. Obergfell, O. Gomonay, J. Sinova, M. Kläui, H. J. Elmers, M. Jourdan, and J. Demsar, Phys. Rev. Lett. **120**, 237201 (2018).

[11]. M. J. Grzybowski, P. Wadley, K. W. Edmonds, R. Beardsley, V. Hills, R. P. Campion, B. L. Gallagher, J. S. Chauhan, V. Novak, T. Jungwirth, F. Maccherozzi, and S. S. Dhesi, Phys. Rev. Lett. **118**, 057701 (2017).

[12]. S. Y. Bodnar, L. Šmejkal, T. Jungwirth, O. Gomonay, A. A. Sapozhnik, H. J. Elmers, M. Kläui, and M. Jourdan, Nat. Commun. **9**, 348 (2018).

[13]. T. Moriyama, K. Oda, T. Ohkochi, M. Kimata, and T. Ono, Sci. Rep. **8**, 14167 (2018).

[14]. X. Z. Chen, R. Zarzuela, J. Zhang, C. Song, X. F. Zhou, G. Y. Shi, F. Li, H. A. Zhou, W. J. Jiang, F. Pan, and Y. Tserkovnyak, Phys. Rev. Lett. **120**, 207204 (2018).

[15]. L. Baldrati, O. Gomonay, A. Ross, M. Filianina, R. Lebrun, R. Ramos, C. Leveille, F. Fuhrmann, T. R. Forrest, F. Maccherozzi, S. Valencia, F. Kronast, E. Saitoh, J. Sinova, and M. Kläui, Phys. Rev. Lett. **123**, 177201 (2019).

[16]. S. DuttaGupta, A. Kurenkov, O. A. Tretiakov, G. Krishnaswamy, G. Sala, V. Krizakova, F. Maccherozzi, and S. Dhesi, Nat. Commun. **11**, 5715 (2020).

[17]. Y. Cheng, S. Yu, M. Zhu, J. Hwang, and F. Yang, Phys. Rev. Lett. **124**, 027202 (2020).

[18]. P. Zhang, J. Finley, T. Safi, and L. Liu, Phys. Rev. Lett. **123**, 247206 (2019).

[19]. X. F. Zhou, X. Z. Chen, J. Zhang, F. Li, G. Y. Shi, Y. M. Sun, M. S. Saleem, Y. F. You, F. Pan, and C. Song, Phys. Rev. Appl. **11**, 054030 (2019).

[20]. X. F. Zhou, J. Zhang, F. Li, X. Z. Chen, G. Y. Shi, Y. Z. Tan, Y. D. Gu, M. S. Saleem, H. Q. Wu, F. Pan, and C. Song, Phys. Rev. Appl. **9**, 054028 (2018).

[21]. L. Q. Liu, O. J. Lee, T. J. Gudmundsen, D. C. Ralph, and R. A. Buhrman, Phys. Rev. Lett. **109**, 096602 (2012).

[22]. H. Yan, Z. X. Feng, P. X. Qin, X. R. Zhou, H. X. Guo, X. N. Wang, H. Y. Chen, X. Zhang, H. J. Wu, C. Jiang, and Z. Liu, Adv. Mater. **32**, 1905603 (2020).

[23]. C. Song, Y. F. You, X. Z. Chen, X. F. Zhou, Y. Y. Wang, and F. Pan, Nanotechnology. **29**, 112001 (2018).

[24]. C. G. Duan, S. S. Jaswal, and E. Y. Tsymbal, Phys. Rev. Lett. **97,** 047201 (2006).

[25]. D. Odkhuu, Phys. Rev. B **96**, 134402 (2017).

[26]. R. O. Cherifi, V. Lvanovskaya, L. C. Phillips, A. Zobelli, I. C. Infante, and E. Iacquet, Nat. Mater. **13**, 345 (2014).

[27]. R. Y. Umetsu, A. Sakuma, and K. Fukamichi, Appl. Phys. Lett. **89**, 052504 (2006).

[28]. J. F. Wang, A. H. Gao, W. Z. Chen, X. D. Zhang, B. Zhou, and Z. Y. Jiang, J. Magn. Magn. Mater. **333**, 93 (2013).

[29]. G. H. Kwei, A. C. Lawson, S. J. L. Billinge, and S. W. Cheong, J. Phys. Chem. A **97**, 2368 (1993)

[30]. P. E. Blöchl, Phys. Rev. B **50**, 17953(1994).

[31]. J. P. Perdew, K. Burke, and M. Ernzerhof, Phys. Rev. Lett. **77**, 3865(1996).

[32]. G. Kresse and J. Hafner, Phys. Rev. B **47**, 558 (1993).

[33]. G. Gerra, A. K. Tagantsev, N. Setter, and K. Parlinski, Phys. Rev. Lett. **96**, 107603 (2006).

[34]. J. Neugebauer and M. Scheffler, Phys. Rev. B **46**, 16067 (1992).

[35]. S. Bornemann, J. Minr, J. Braun, D. Kdderitzsch, and H. Ebert, Solid State Commun. **152**, 85 (2012).

[36]. Y. R. Su, M. Y. Li, J. Zhang, J. Hong, and L. You, J. Magn. Magn. Mater. **505**, 166758 (2020).

[37]. Y. R. Su, J. Zhang, J. Hong, and L. You, J. Phys. Condens. Matter. **32**, 454001 (2020).

[38]. J. Kim, K. W. Kim, B. Kim, C. J. Kang, D. Shin, S. H. Lee, B. C. Min, and N. Park, Nano Lett. **20**, 929 (2020).

[39]. J. L. Lado and J. Fernández-Rossier, 2D Mater. **4**, 35002 (2017).

[40]. D. Odkhuu, Sci. Rep. **8**, 6900 (2018).

[41]. P. H. Chang, W. Fang, T. Ozaki, and K. D. Belashchenko, Phys. Rev. Mater **5**, 054406 (2021).

[42]. D. S. Wang, R. Q. Wu, and A. J. Freeman, Phys. Rev. B **47**, 14932 (1993).

[43]. P. Bruno, Phys. Rev. B **39**, 865 (1989).

[44]. G. Van der Laan, J. Phys. Condens. Matter **10**, 3239 (1998).

[45]. J. Zhang, P. V. Lukashev, S. S. Jaswal, and E. Y. Tsymbal, Phys. Rev. B **96**, 014435(2017).

[46]. L. Szunyogh, B. Újfalussy, and P. Weinberger, Phys. Rev. B **51**, 9552 (1995).

[47]. M. Woinska, J. Szczytko, A. Majhofer, J. Gosk, K. Dziatkowski, and A. Twardowski, Phys.Rev. B **88**, 144421 (2013).